\documentclass{cs23proc}

\usepackage{kantlipsum}

\editors{Takeru Suzuki and the Cool Stars 23 Organizing Team}
\publisher{Zenodo}
\conference{The 23th Cambridge Workshop on Cool Stars, Stellar Systems, and the Sun (Cool Stars 23)}
\conferencedate{2026}

\title{Observational Insights into Post Main-Sequence Rotation and Magnetism: 20 Years of Science, from the Subgiant to the White Dwarf Phase.}
\author{Emily J. Hatt $^{1,2}$}

\affiliation{$^{1}$ Institute of Science and Technology Austria (ISTA), Am Campus 1, Klosterneuburg, Austria  \\
$^{2}$ School of Physics and Astronomy, University of Birmingham, Birmingham B15 2TT, United Kingdom \\ }

\shorttitle{Observing post main-sequence rotation and magentism in stars}
\shortauthors{Emily J. Hatt}

\abs{Rotation and magnetism are important phenomena to consider when trying to understand stars. By shaping the structure and chemical mixing in the interior of stars, they can induce significant changes in both their fundamental and observable properties. Despite this, our theoretical picture of how rotation and magnetism evolve throughout the stellar lifecycle remains incomplete. Gaps in our understanding are particularly stark for evolved stars, where slow rotation rates and weak magnetic fields make observational studies difficult. In response to this issue, the last 20 years have seen instrumentation to probe magnetism and rotation progressing in leaps and bounds. Through these new tools we have found that evolved stars rotate in ways that we cannot predict and have magnetic fields with properties that we did not expect. In the following, I will provide a brief overview of some recent observational results and their implications for our understanding of stars. I will focus on advances in asteroseismology and high-precision spectropolarimetry, as these techniques have advanced significantly over the last few decades.  In keeping with the theme of Cool Stars, I'll cover stars with low to intermediate masses and their evolution from the subgiant to the white dwarf phase.}

\begin{document}

\maketitle

\section{Introduction}
\textit{Have you considered the impact of rotation and magnetism on your work?} This is a question we often hear at stellar physics conferences, and for good reason. While we know rotation and magnetism are at work in stars, we don't have a complete understanding of how they operate across the entire interior or stellar lifecycle. It is clear that the two phenomena can be closely linked, this is certainly the case for the magnetic field we observe in the outer layers of the Sun, which manifests as features like sun spots\footnote{The human study of sun spots dates back to at least 800 BCE, when astronomers in China note the appearance of spots in the book \textit{I Ching} -- this was $\approx$ 45 years before the founding of Rome.}. The process that sustains this field is known as the solar-type dynamo and while there is still a lot to be learnt about how it works, it is known to depend on the joint action of rotation and convection \citep[see, for example, the review of][]{2023SSRv..219...35C}. As you will see in the later discussions of this review, the fields generated by this dynamo are often parametrized by the ratio of the rotational to convective forces. This is known as the Rossby number, or Ro, and can be rewritten as a ratio of the relevant timescales -- that is the rotational period (P$_{\mathrm{rot}}$) over the convective turnover time\footnote{This is a timescale associated with mixing length theory, and scales with the ratio of the mixing length to the turbulent velocity.} \citep[$\tau_{\mathrm{conv}}$, ][]{1984ApJ...279..763N}.

Observations of magnetic fields generated by solar-type dynamos in stars other than the Sun have been made for many years, confirming that this mechanism can operate efficiently so long as Ro is less than $\approx$1  \citep[see, for example][]{2001MNRAS.326..877M, 2018A&A...618A..48M, 2025ApJ...995...32B}. While this is generally the case for low to intermediate mass stars on the main sequence \citep[e.g.][]{2021A&A...652L...2C}, the increase in P$_{\mathrm{rot}}$ incurred as they age and join the subgiant branch\footnote{This is largely the result of stellar-wind driven spin down on the main sequence \citep[e.g.][]{1967ApJ...148..217W, 1967ApJ...150..551K, 1972ApJ...171..565S, 2015A&A...577A..28J}, and the expansion of the outer layers on the subgiant branch.} means that we expect solar-type dynamos to become less efficient as we move toward evolved stars.

While it has been a long-held expectation that both rotation and magnetism are not dominant mechanisms in the outer layers of cool evolved stars, stringent evaluation of this belief against a large body of observational evidence is a fairly recent development. This has primarily been the result of the observational challenges associated with exploring either phenomenon in stars that have evolved off of the main sequence. Consider, for example, the recovery of stellar rotation rates through spectroscopy. Although we don't have an absolute theory for the evolution of stellar rotation, we do observe that evolved stars generally have outer layers that rotate slowly \citep[typical envelope rotation rates on the red giant branch are < 1kms$^{-1}$][]{1981ApJ...251..155G, 1996A&A...314..499D, 2024A&A...688A.184L}. Additionally, evolved stars exhibit large macroturbulent velocities, often having the same magnitude as the rotational velocities. Spectroscopic recovery thus requires searching for a small shift to a line which has been significantly broadened by macroturbulence. Accurate and precise recovery thus requires high precision spectrometers and careful data analysis.

Similar challenges exist for the recovery of the magnetic field properties of evolved stars. A direct method to observationally study stellar magnetism is by monitoring its impact on the polarisation of starlight (I'll cover the basics of this technique in \ref{sec:spectro}). The degree to which light is polarised, and thus the strength of the observational signal, is strongly dependent on the magnetic field strength.
It appears to be true that the magnetic fields at the surfaces of low to intermediate mass evolved stars are much weaker than those observed on the main sequence, when they are detectable at all \citep[e.g.][]{2017A&A...605A.102C}. Recovering detailed measurements of the magnetic fields at the surfaces of evolved stars thus requires very high precision instrumentation.

So far this discussion has been limited to the outer layers of stars. Given that the structure and dominant process vary throughout stellar bodies -- often significantly so -- it is important to also consider what is happening deep in the interior. Let us naively assume all we need for the solar-type dynamo is convection. As in the Sun, intermediate mass (0.3M$_{\odot}$ $\leq$ M $\leq$ 2M$_{\odot}$) stars on the main sequence have outer convection zones. However, most of the stellar interior is not convective but radiative, so we cannot assume that the field regenerated in the outer layers is the same in the deep interior. Additionally, main sequence stars with masses $\geq$ 1.1M$_{\odot}$ have convective cores. Theoretically, it is possible for these stars to support two solar-type dynamos, one in the outer convection zone, and one in the core convection zone \citep[see for example,][]{2005ApJ...629..461B}.

In evolved stars the difference between core and surface properties becomes even more extreme -- as the cores contract and the surfaces expand. Considering rotation, if angular momentum is conserved locally this change would result in steep rotational gradients across the star. Conversely, if angular momentum is conserved globally (that is there is some mechanism coupling the core to the surface), the whole star should spin down. This changing rotation, in combination with the changing conditions in the deepening outer convection zone should have implications for the properties of a solar-type dynamo.

Motivated by the need to better understand rotation and magnetism both as it varies in stellar interiors, and across evolutionary timescales, observational instrumentation has greatly advanced in the last 20 years. This includes the development and installation of high precision spectrographs and spectropolarimeters, which I will discuss in section \ref{sec:spectro}. Measurements from such instruments have revolutionised our understanding of the properties of the outer layers of stars, but they can't tell us what is happening deep in the interior. Fortunately, the last 20 years have also seen a renaissance in the field of asteroseismology -- the study of stellar pulsations. Unlike the more `classical' probes of stellar properties, asteroseismic data is directly sensitive to the interior, where it can be used to recover both rotational and magnetic properties. I will give a brief overview of asteroseismology and its application in section \ref{sec:astero-theory}.

\section{Methods}
\subsection{Spectropolarimetry}\label{sec:spectro}
Polarimetry is the study of the polarisation of light. When atoms are placed in a magnetic field, they experience the Zeeman effect -- the splitting of energy levels. The light emitted from transitions between the different levels in Zeeman split components is differently polarised depending on the field strength and orientation. Thus by studying the net polarisation of light emitted by a star -- polarimetry -- or the polarisation as a function of wavelength -- spectropolarimetry -- we can infer something about the properties of the magnetic field in the photosphere. 

A full description of the properties of the polarised light can be obtained by measuring four parameters -- one parameter to capture the total polarised light, one capturing how much is linearly polarised horizontally/vertically, one for the linear polarisation at $\pm$45$^{\circ}$, and one for the circular polarisation. These four parameters are known as the Stokes parameters, and labelled I, Q, U and V. Spectropolarimetry measures these parameters as a function of wavelength, while polarimetry measures the integral over the disk.

By studying all four stokes parameters as a function of wavelength, one can reconstruct topological maps for the distribution of the magnetic field over the stellar surface. However, the field strengths we expect in most low to intermediate mass stars induce very small amounts of polarisation: The largest signature is in stokes V, where even kG fields will only result in percent changes in the continuum polarisation \citep{Kochukhov2016}. Spectropolarimetric recovery of the field strengths typical of evolved stars \citep[a few Gauss,][]{2014IAUS..302..373K} thus requires very high precision instrumentation. 

In recent years a number of high precision spectropolarimeters and polarimeters have been installed across the world. While writing this review I came across the following frequently;
\begin{itemize}
    \item ESPaDOnS \citep{10.1117/12.458230} - installed in 2003 at the CFHT.
    \item DIPOL-UF \citep{2020arXiv201102129P} - installed in 2020 at the NOT.
    \item PEPSI \citep{2015AN....336..324S} - installed in 2014 at the LBT.
    \item SPHERE-ZIMPOL \citep{2019A&A...631A.155B} - installed in 2014 at the VLT.
\end{itemize}

\subsection{Asteroseismology} \label{sec:astero-theory}
Asteroseismology is the study of stellar pulsations. We generally classify pulsating stars into one of two groups; classical pulsators and solar-like oscillators. In classical pulsators the oscillations are driven by what's known as the kappa mechanism, a type of heat engine functioning in regions in the star where the opacity increases with increasing temperature. Such regions are supported in relatively high mass, hot stars and so I won't be discussing them in this review.

Cool stars, on the other hand, can support solar-like oscillations, which are driven and damped by the turbulent motion associated with convection. This process drives sound waves that propagate through the stellar body, where they are refracted and reflected until they interfere, resulting in trapped pressure mode oscillations. This class of pulsator will feature heavily in the following.

Solar-like oscillations cause periodic expansions and contractions of the star at supported overtone frequencies, which we describe mathematically using spherical harmonics. Accordingly, overtones can be referred to by three indexes -- the radial order ($n$), angular degree ($\ell$) and azimuthal order ($m$). The process of observing stars involves integrating the oscillation signal over the observable disk. The resulting geometric cancellation means that only modes with $\ell \leq 3$ are measurable. In the following you will read references to radial modes, those are modes with $\ell$ = 0, and non-radial modes, which are modes with $\ell = 1, 2$ or 3.

\begin{figure}
    \centering
    \includegraphics[width=0.9\linewidth]{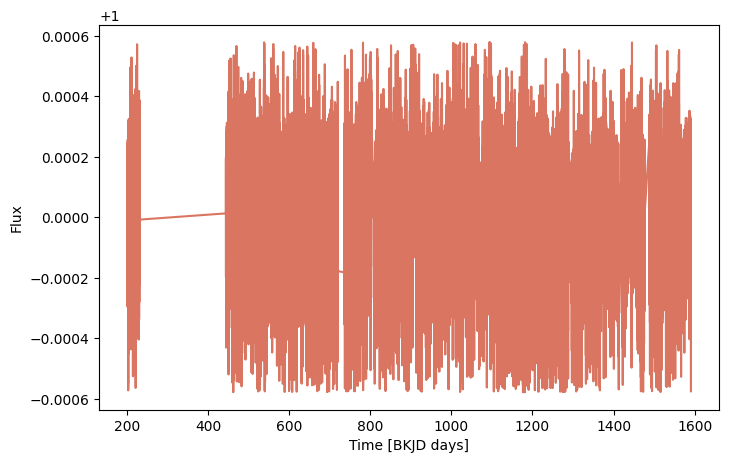}
    \caption{Lightcurve of a solar-like oscillator observed by \textit{Kepler}.}
    \label{fig:lightcurve}
\end{figure}

\begin{figure}
    \centering
    \includegraphics[width=0.98\linewidth]{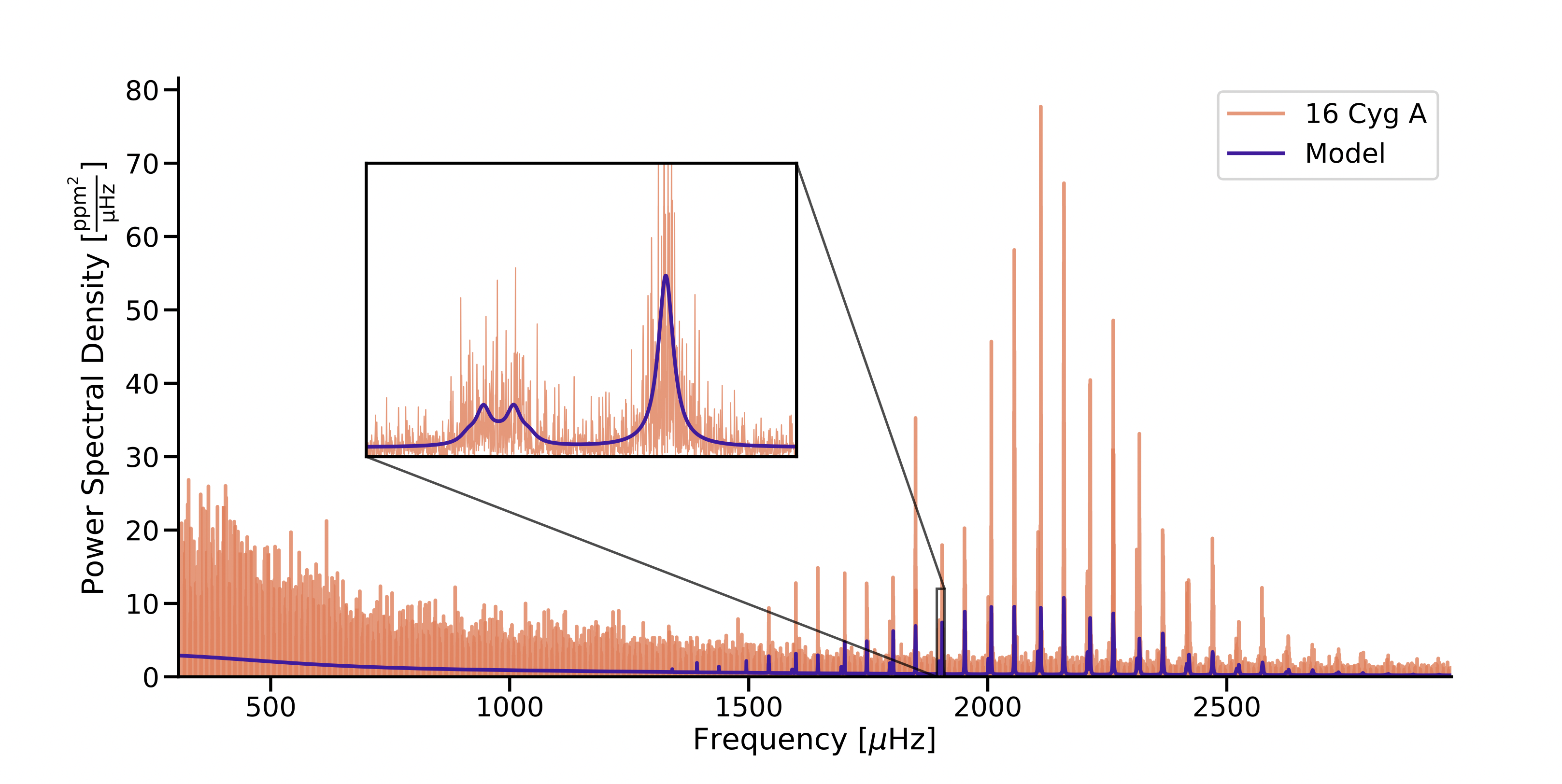}
    \caption{Supplementary figure 2 in \citet{2021NatAs...5..707H} showing the frequency spectrum of a solar-like oscillator observed by \textit{Kepler}. Inset axes focus on a pair of overtone frequencies, the overtone on the left has been impacted by rotation.}
    \label{fig:spectrum}
\end{figure}

The motion of the surface caused by solar-like pulsations means that we can identify them as periodicities in spectroscopic measurements. Additionally, the fluctuations in temperature as a result of the compression and rarefaction of the gas cause the luminosity to periodically change, and so they are also recoverable in photometry. Regardless of whether we are looking in spectroscopy or photometry, the stochastic nature and small amplitudes of the pulsations means that solar-like oscillations are hard to identify by eye in the time domain -- see the lightcurve in figure \ref{fig:lightcurve}. Visually, the signature is far more striking in the frequency domain. This is exemplified in figure \ref{fig:spectrum} taken from \citet{2021NatAs...5..707H}, which shows the power spectral density as a function of frequency for a solar-like oscillator observed by \textit{Kepler}. There-in, the pulsations appear as sharp Lorentzian peaks centred on the overtone frequencies of the star's pulsations. The amplitudes of these peaks are modulated by a roughly Gaussian envelope, centred on a frequency which scales with the surface gravity (logg) and inverse square root of the effective temperature -- i.e. more evolved stars pulsate at lower frequencies than their main sequence counterparts. The frequencies and properties of the line profiles for individual overtones depend on the properties of the gas within which they propagate. These properties include rotation and magnetism, as I will discuss shortly. 

For solar-like oscillators on the main sequence, the modes are primarily sensitive to properties in the outer convection zone where they are excited. However, these convection-driven modes aren't the only class of oscillation solar-likes support. Deep in the stellar interior, near the stellar core, the strong density stratification allows the gas to support gravity modes. As they are buried close to the core, these modes have a negligible impact on the star's surface and are therefore not directly measurable. Additionally, for stars on the main sequence, the gravity-mode frequencies are much lower than those of the pressure modes, and the two classes of pulsation don't interact. However, as stars evolve off of the main sequence, the frequencies of the pressure modes decrease, while the frequencies of the gravity modes increase. The two classes of mode approach each other in frequency, and (once they are close enough) the non-radial pressure modes couple to the gravity modes. The resultant modes, known as \textit{mixed modes}, are sensitive both in the outer envelope and in the core. Therefore, in subgiants and red giants, non-radial\footnote{These modes are non-radial only, as gravity modes cannot be purely radial.} solar-like pulsations carry information about what is happening both at the surface and near the core.

Gravity mode, pressure mode and mixed mode pulsations can be impacted by rotation and magnetism. Both phenomena can induce observable shifts to mode frequencies. In the case of rotation, this shift is primarily the result of advection and can be thought of in terms of the Doppler effect -- modes travelling with the rotation are shifted upward in frequency, those against shift downward. For a magnetic field the shift is induced by the introduction of Lorentz force into the oscillation equations -- this can be compared to the Zeeman effect. Additionally, it has been shown that strong magnetic fields in the cores of evolved stars can suppress the amplitudes of mixed modes. Therefore, works looking at internal rotation and/or magnetism search for perturbations to the frequency pattern or `missing' mixed modes (indicating magnetic mode suppression). 

Asteroseismology has become increasingly popular in the last 20 years, due in large part to the amount of data available. Analysis requires high precision measurements, made at high cadence, and over periods of at least several weeks. Gaps in data can also cause issues for asteroseismology, as they induce alias' in the power spectrum that can complicate the recovery of pulsations, so missions designed to `stop and stare' are preferred. These requirements have been met by a number of high-precision space-based photometric telescopes that have launched over the last 20 years. These include;
\begin{itemize}
    \item CoRoT \citep{2009A&A...506..411A} - launched in 2006.
    \item \textit{Kepler}/K2 \citep{2010Sci...327..977B, 2014PASP..126..398H} - launched in 2009.
    \item TESS \citep{2015JATIS...1a4003R} - launched in 2018.
    \item (upcoming) The Nancy Grace Roman Telescope \citep[][, previously known as WFIRST]{2015arXiv150303757S, 2019ApJS..241....3P} - due to launch August 2026.
    \item (upcoming) PLATO \citep{2014ExA....38..249R, 2025ExA....59...26R}- due to launch in 2027.
\end{itemize}

\section{Results}
So the important question is; what have these techniques and new instruments told us? In the following sections I will cover some important results from the last $\approx$20 years of science, from the subgiant phase through to the white dwarfs. In keeping with the Cool Stars conference theme, to which this talk was given, I'll be primarily discussing results in low-intermediate mass stars. To keep things parsable, I'll discuss rotation and magnetism separately, but given the degree of dependence between the two, when one is discussed you may assume it has implications for the other. 

\subsection{Subgiants}
\subsubsection{Rotation}\label{sec:SGB-rot}
\begin{figure}
    \centering
    \includegraphics[width=0.95\linewidth]{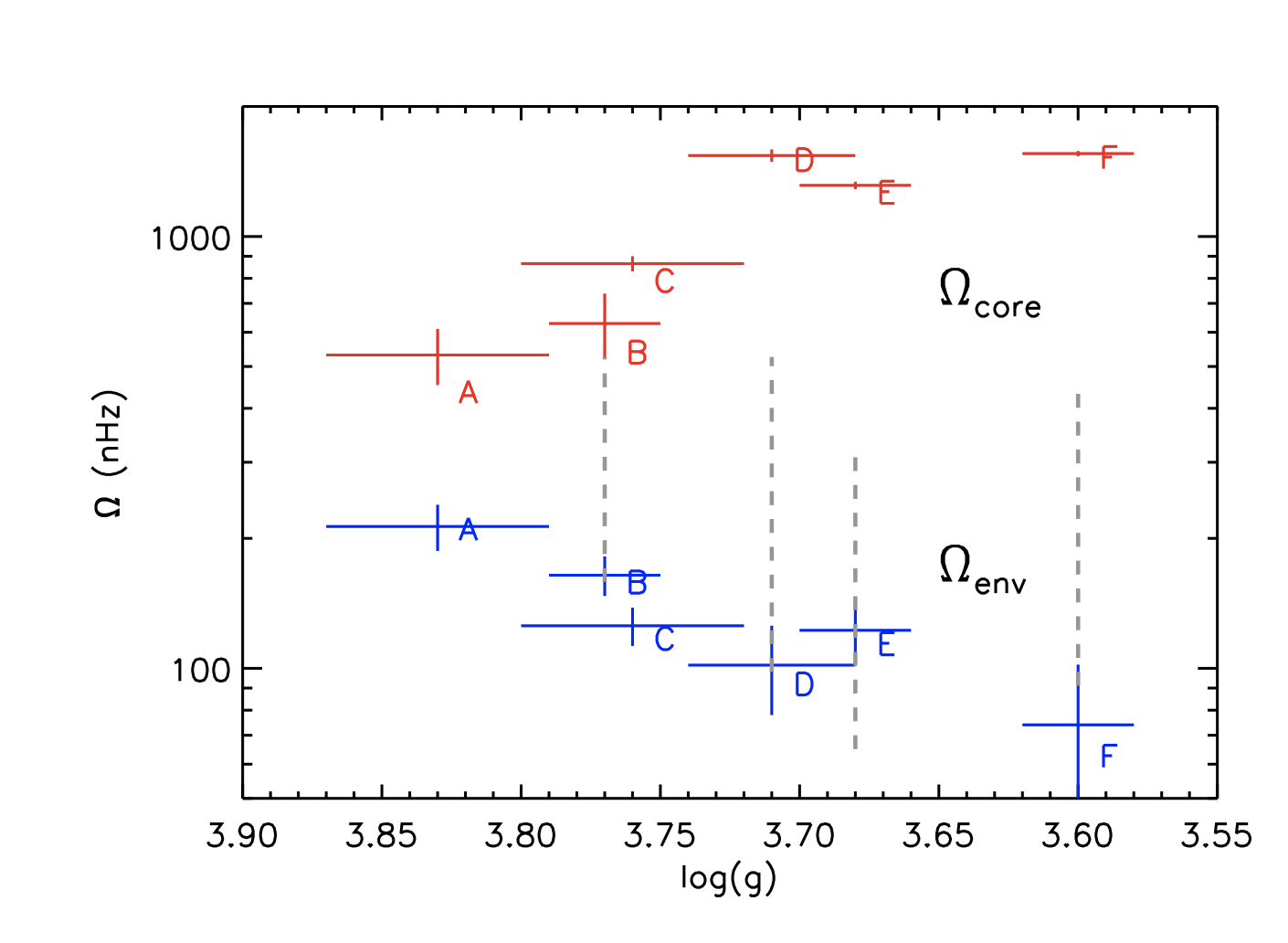}
    \caption{Figure 13 from \citet{2014A&A...564A..27D}: Angular rotation rates for the cores and envelopes of six subgiants observed by \textit{Kepler}. Core rotation is in red, envelope rotation is in blue.}
    \label{fig:subgiant-internal-rotation}
\end{figure}
As mentioned in section \ref{sec:astero-theory}, solar-like oscillations in subgiants and more evolved stars can be used to probe rotation rates both in the near core layers and in the outer convection zone. In 2014, \citet{2014A&A...564A..27D} capitalised on this, using \textit{Kepler} data to recover the first measurements of both core rotation and envelope rotation rates in six subgiants. The results of this recovery can be seen in figure 13 of that work, included here as figure \ref{fig:subgiant-internal-rotation}. In this figure the surface gravity decreases to the right, so we can approximate this as stars moving from left to right on this plot as they evolve. 
There are two things I'd like to note about the data in this figure:
\begin{enumerate}
    \item In all six stars the cores are rotating more rapidly than the envelopes.
    \item Core rotation rates increase as we move toward the more evolved stars, while the envelope rotation rates decrease -- in other words, the rotational shear increases with evolution.
\end{enumerate}
This was the first direct observational evidence we had for local angular momentum conservation in subgiants. That is, as these stars evolve their cores contract, should angular momentum be conserved in just the core separate to the rest of the star, this should imply the core spins up. On the other hand, the envelopes are expanding, so again in keeping with local conservation of angular momentum, we'd expect the outer layers to spin down. While this work is, at time of writing, over ten years old, it still remains the largest catalogue of internal rotation rates in subgiant stars. 

\subsubsection{Magnetism} \label{sec:subgiant-mag}
We know that solar-type dynamos act in the outer convection zones of low mass stars on the main sequence. We also believe that the efficiency of this mechanism depends on the ratio of the rotational period to the convective turnover time. We've seen that the envelopes of subgiants spin down as they evolve. Therefore, we might expect that (compared to their main sequence counterparts) magnetism doesn't play a big role in the outer layers of low to intermediate mass subgiants and more evolved stars. 

\begin{figure}
    \centering
    \includegraphics[width=0.95\linewidth]{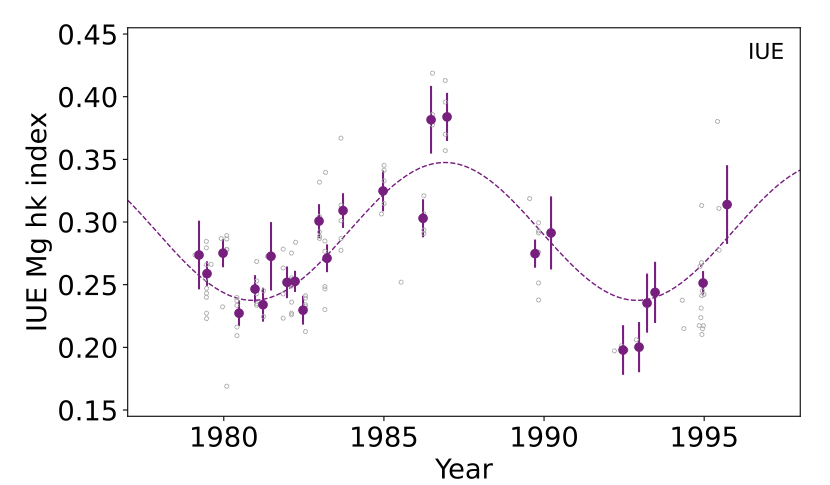}
    \caption{Panel from figure 1 in \citep{2024ApJ...974...31M} showing an activity indicator for the subgiant $\beta$ Hydri as a function of time. The dashed line shows a sinusoidal variation at the period inferred for the activity cycle in this star.}
    \label{fig:beta-hydri}
\end{figure}

However, \citet{2024ApJ...974...31M} studied a subgiant ($\beta$ Hydri) which doesn't appear to obey this trend. $\beta$ Hydri has a very similar mass and chemical composition to the Sun, but is older ($\approx$ 6.5 Gyrs) and has evolved off of the main sequence. Therefore, we would expect that it would have an activity cycle with a longer period than the cycle in the Sun. Instead \citet{2024ApJ...974...31M} reports an activity cycle in the star that has a period of $\approx$12 yrs, on par with the dominant cycle in the Sun. The work collected asteroseismic measurements, chromospheric activity indicators and a photometric rotation rate for $\beta$ Hydri, using these measurements to constrain rotation evolution models. Some of the data used for recovering the cycle period is shown in figure \ref{fig:beta-hydri} which presents the Mg HK index (a proxy for activity) measured by the International Ultraviolet Explorer as a function of time. 

\citet{2024ApJ...974...31M} finds that in $\beta$ Hydri the expansion of the outer layers increases the convective turnover time such that, even with the decreasing surface rotation rates, Ro drops below 1. It is argued that this results in the regeneration of an efficient solar type dynamo. This phase is found to be transient as after a period of $\approx$ 1 Gyr the rotation rate is slow enough that again the dynamo process is expected to become inefficient. 

\subsection{Red giants}
\subsubsection{Rotation} \label{sec:RGB-rot}
\begin{figure}
    \centering
    \includegraphics[width=0.95\linewidth]{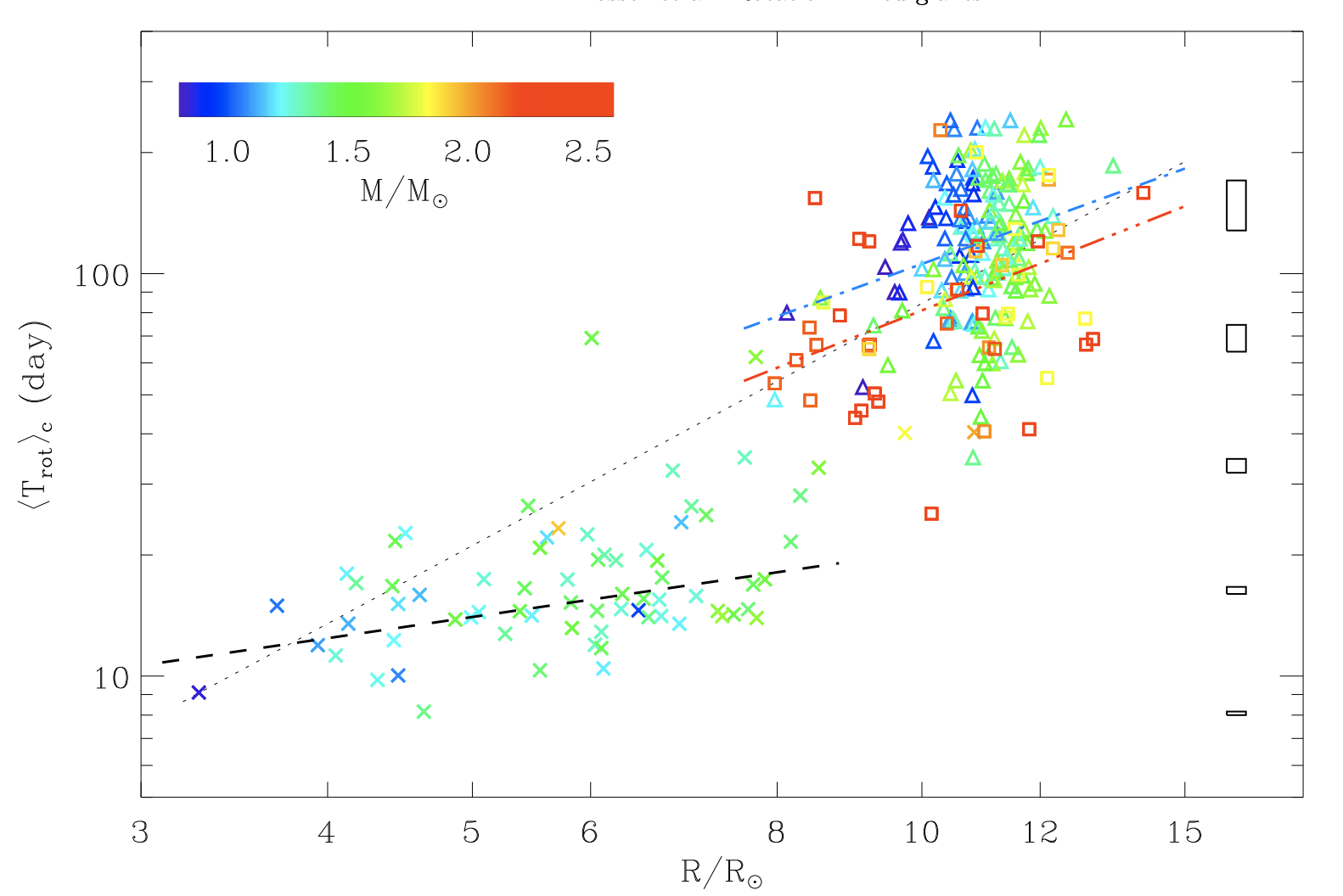}
    \caption{Figure 9 from \citet{2012A&A...548A..10M}: Core rotational period as a function of stellar radius (R) for $\approx$ 300 red giant and clump stars. Coloured scatter points are the measured rotational periods, coloured by stellar mass. Crosses mark stars on the red giant branch, triangles are clump stars and squares are secondary clump stars. The dashed, dot dashed and triple dot dashed lines are fits to the data for the stars of different evolutionary state, the dotted line is the expectation should the core rotation vary as $\propto$ R$^2$.}
    \label{fig:Mosser-RGB-rot}
\end{figure}

Two years prior to the measurements discussed in section \ref{sec:SGB-rot}, the first measurements of internal rotation rates from asteroseismology were made. These were published in \citet{2012Natur.481...55B} and \citet{2012A&A...548A..10M}, which catalogued the core rotation rates in over 300 red giants and red clump stars. The results can be summarised by figure 9 in \citet{2012A&A...548A..10M}, which is printed here as figure \ref{fig:Mosser-RGB-rot}. There-in the stellar radius increases to the right such that we can read stars as evolving from the left to the right.
Here I'll note another two points:
\begin{enumerate}
    \item The measured core rotation periods are increasing as we move to more evolved stars. That is, the cores are now spinning down rather than spinning up.
    \item The dotted line marks the evolution of the core rotation if it were tracing the expansion of the outer envelope. This is clearly not a good fit to the data (note the y axis is in log scale).
\end{enumerate}
This scenario is very different to what was observed in the subgiants. Rather than the cores spinning up as they contract, they have rotation rates which appear to be slowing down. We cannot have the same flavour of local angular momentum conservation as we saw in the subgiants. The fact that the rotation rates aren't varying $\propto$R$^2$ also tells us that we cannot have complete coupling from the core to the surface. So \textit{what is going on?}

At the time that these rotation rates were published, none of our models for stellar rotation could account for the observed trend with radius. In addition to not reproducing the spin down, models predicted core rotation rates that were orders of magnitude more rapid than those shown in figure \ref{fig:Mosser-RGB-rot}. This issue was attributed to the models missing some mechanism that transports angular momentum from the cores of stars to their surfaces -- a flaw subsequently donned the \textit{missing angular momentum transport problem}. This problem remains unsolved up to time of writing, but there are several established theories posing mechanisms that could be at work. These include; internal gravity waves \citep[e.g.][]{2014ApJ...796...17F}, mixed modes \citep[e.g.][]{2015A&A...579A..30B} and a magnetic field coupling the core to the surface \citep[e.g.][]{2019MNRAS.485.3661F}.

\subsubsection{Magnetism}\label{sec:RGB-mag}

\begin{figure}
    \centering
    \includegraphics[width=0.99\linewidth]{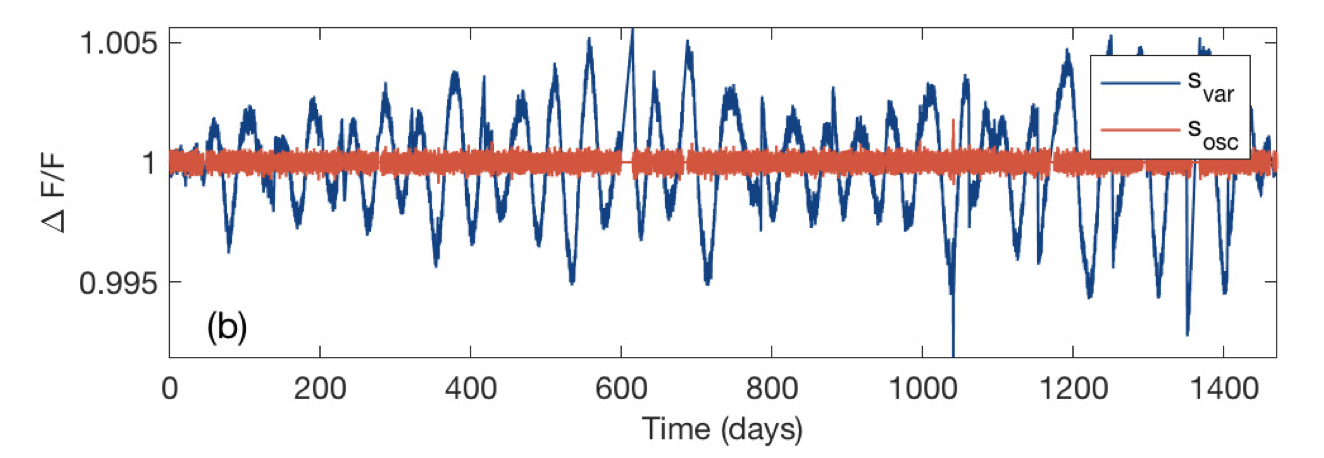}
    \caption{Excerpt from figure 5 in \citet{2020A&A...639A..63G} showing the lightcurve of an active red giant as observed by \textit{Kepler}. In blue is the lightcurve including both activity (starspots) and oscillations, in red everything but the oscillations have been removed.}
    \label{fig:active-rg-lightcurve}
\end{figure}
As we discussed in section \ref{sec:subgiant-mag}, we expect that as stars evolve and their surfaces spin down, the solar like dynamo process should become less and less efficient. One might expect then that we don't observe the signature of magnetic fields in the outer layers of red giants. To evaluate the prevalence of detectable magnetic fields in the outer layers of such stars \citet{2020A&A...639A..63G} collected lightcurves for $\approx$ 4500 giants observed by \textit{Kepler}. Therein a signature of rotational modulation consistent with star spots (and thus magnetism) was found in roughly 8\% of the sample. Figure \ref{fig:active-rg-lightcurve} is excerpt of a figure from the work, showing an example lightcurve for one of the active giants.

\begin{figure}
    \centering
    \includegraphics[width=0.95\linewidth]{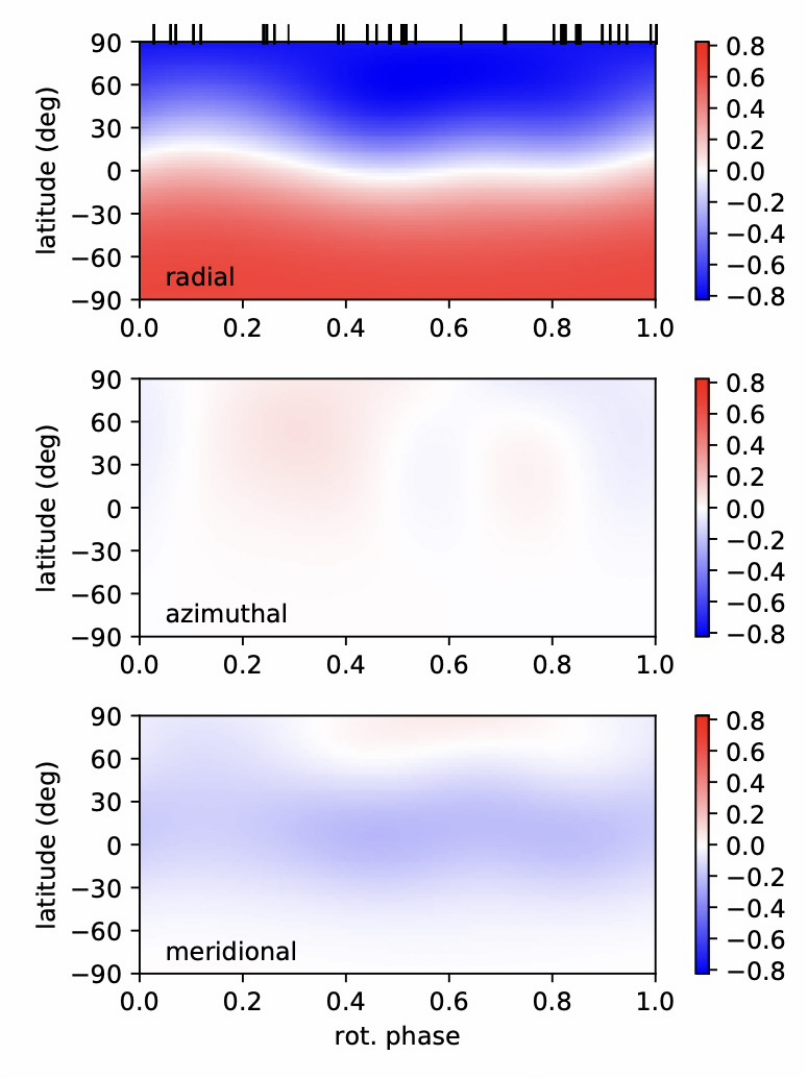}
    \caption{Figure 7 from \citet{2021A&A...646A.130A} showing a reconstruction of the magnetic field observed in the red giant Pollux. The colour bar is the average field strength in Gauss.}
    \label{fig:Pollux}
\end{figure}

While this work aimed to provide a statistical view of the prevalence of activity in red giants, advances in spectropolarimetry have started to enable more detailed investigations of the magnetic properties of specific targets. An example of this is the work of \citet{2021A&A...646A.130A}, which used spectropolarimetric measurements made with the ESPaDOnS and Narval instruments to infer the magnetic properties of Pollux -- a K0III type red giant. This work found a weak but measurable field, with a longitudinal field strength that varies periodically, having a mean field strength of 0.4G. The work also produced the first topological map of the field on Pollux, which is included here in figure \ref{fig:Pollux}, showing the field's dipolar distribution.

The aforementioned works proved that magnetic fields are detectable in the outer layers of at least some red giants. However, these works do not discuss the presence of magnetism in the deep interior. To probe this observationally, we must turn to asteroseismology, as discussed in section \ref{sec:astero-theory}. The first observational work to that end was that of \citet{2016Natur.529..364S}, which searched for magnetic mode suppression in a sample of over 3600 red giants observed by \textit{Kepler}. There-in they identified that about 20\% of the sample showed a magnetic signature, implying minimum core field strengths from 10s kG to $>$1 MG. 

The method used in \citet{2016Natur.529..364S} can only recover constraints on the minimum field strength in a star with suppressed modes. On the other hand, measurements of the perturbations to mode frequencies could enable us to recover both the average field strength directly, and put some constraints on how that field is distributed spatially. The first of such measurements was made in \citet{2022Natur.610...43L}, which measured these effects in three red giants observed by \textit{Kepler}. The measured signatures in these stars were identified as consistent with average radial core field strengths of 30 kG - 100 kG.

\begin{figure}
    \centering
    \includegraphics[width=0.98\linewidth]{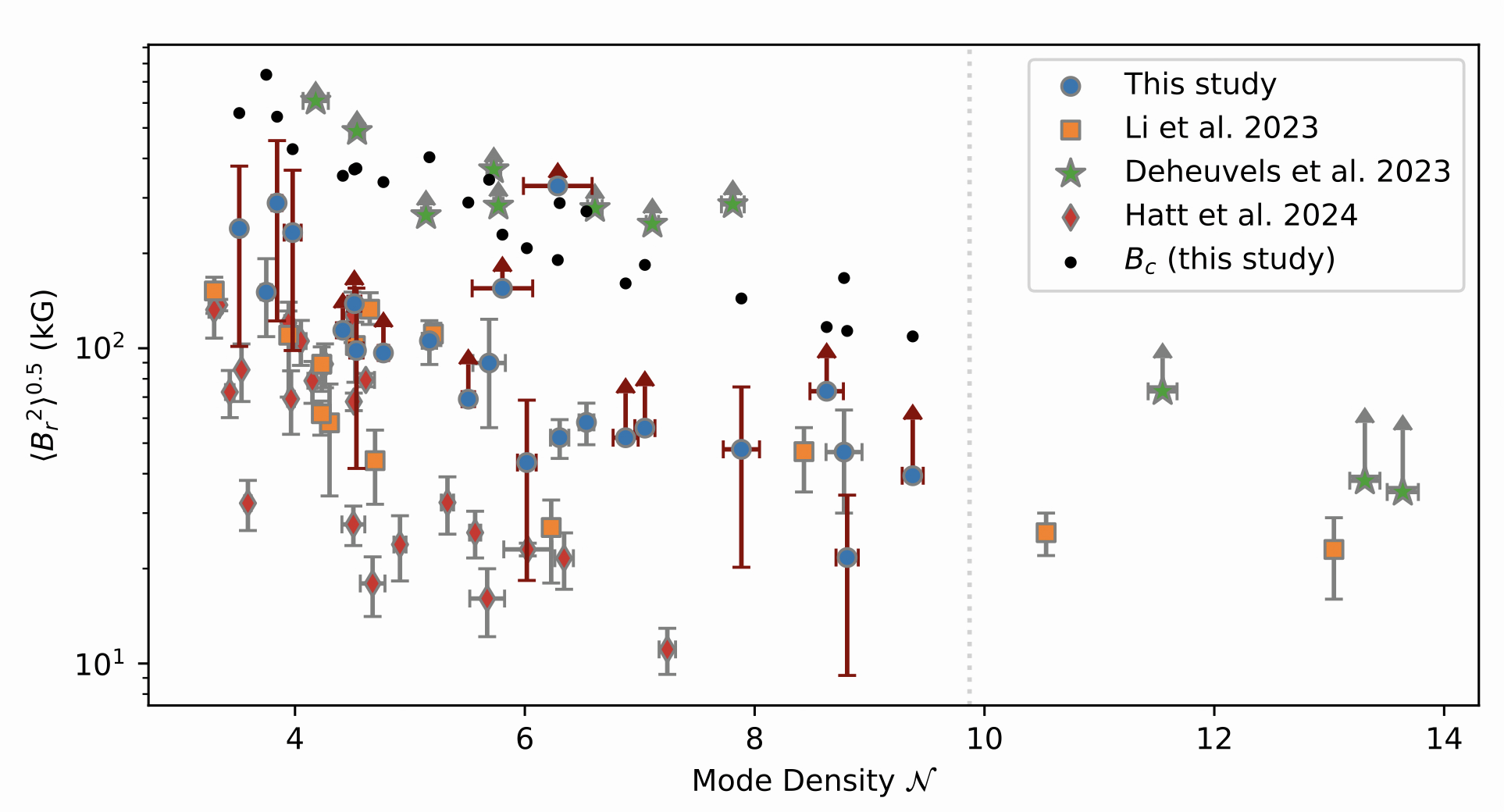}
    \caption{Figure 12 from \citet{2026A&A...707A.366V} showing the average field strength in the cores of $\approx$70 red giants as a function of an asteroseismic parameter, $\mathcal{N}$, which traces the evolution. Stars evolve towards the right on this plot.}
    \label{fig:B-versus-N}
\end{figure}

Since those first measurements were made, a number of works have followed applying different techniques to recover these measurements in a larger sample of stars \citep[e.g.][]{2023A&A...670L..16D, 2023A&A...680A..26L, 2024MNRAS.534.1060H}. At time of writing, significant detections of core magnetic fields have been made for $\approx$ 70 stars, which are summarised by figure 12 of \citet{2026A&A...707A.366V} shown here as figure \ref{fig:B-versus-N}. Here it appears that the average field strengths decrease with increasing evolution. However, it should be noted that the critical field strength above which modes are suppressed decreases with evolution. This trend is thus likely reflecting the fact that for more evolved stars, modes start to drop out of the spectrum at lower field strengths, and thus we cannot measure mode perturbations. 

\subsection{AGB stars}
\subsubsection{Rotation}
\begin{figure*}
    \centering
    \includegraphics[width=0.98\linewidth]{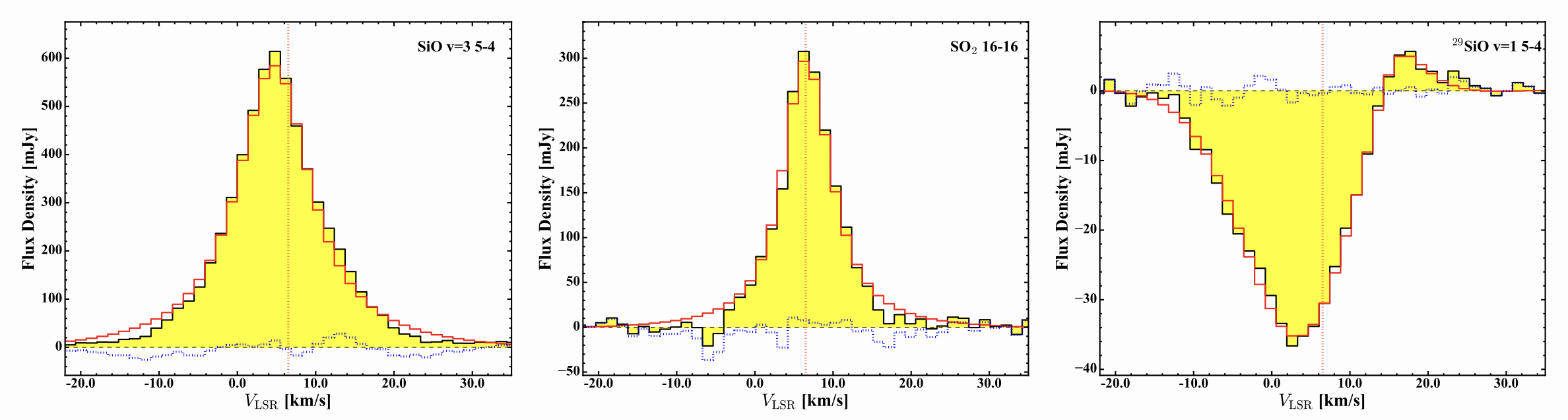}
    \caption{Figure A.1 from \citet{2018A&A...613L...4V} showing the emission and absorption lines measured using ALMA and used to constrained the rotation rate in the envelope of the AGB star R Doradus.}
    \label{fig:R-Dor-rot}
\end{figure*}

According to the expansion of the envelope on the AGB, and given the already slow rotation rates observed in the surfaces of most red giants, I expected prior to writing this review that rotation would not be measurable in AGB stars. While it does appear to be true that the recovery of rotation in AGB stars is observationally challenging, it has been done. This is exemplified by the work of \citet{2018A&A...613L...4V}, which recovered the envelope rotation rate of the AGB star R Doradus. Figure \ref{fig:R-Dor-rot} is taken from figure A.1 in that work, and shows the lines used to constrain the rotation rate, which was found to be $|\mathrm{vsin}i| = 1.0 \pm 0.1$ kms$^{-1}$. According to that work, this is much more rapid than would be expected for traditional single star evolution, which would have the rotation rate on the order of 0.1 kms$^{-1}$, pointing to some non standard spin-up mechanism. 

\subsubsection{Magnetism}
Given the correlation between rotation and magnetism, and my expectation for the rotation rates in AGB stars, I had also expected these stars to be not detectably magnetic. Perhaps I shouldn't be surprised given my personal revelation regarding measurements of rotation, but I discovered that AGB magnetism is a very active field of research. Measurements of magnetic fields in AGB stars are reported in a number of works including the catalogue of \citet{2014IAUS..302..373K}. There-in, spectropolarimetric detections of magnetic fields were made in 8 AGB stars from a sample of 13. These fields had maximum longitudinal field strengths of a few Gauss. 

\subsection{White Dwarfs}
\subsubsection{Rotation}
White dwarfs mark the final phase in the lifecycle of most low-intermediate mass stars. After the outer layers of the star are stripped away on the AGB, what remains is the dense, degenerate stellar core. Given their compact nature, they were historically expected to be rapid rotators. The first suggestion that this might not be the case was made in the 90s, when it was discovered that rotation rates of samples of white dwarfs were slow enough to be consistent with zero rotation given the data quality at the time \citep[vsin$i \lessapprox$ 15km$s^{-1}$][]{1998A&A...338..612K}. 

\begin{figure}
    \centering
    \includegraphics[width=0.9\linewidth]{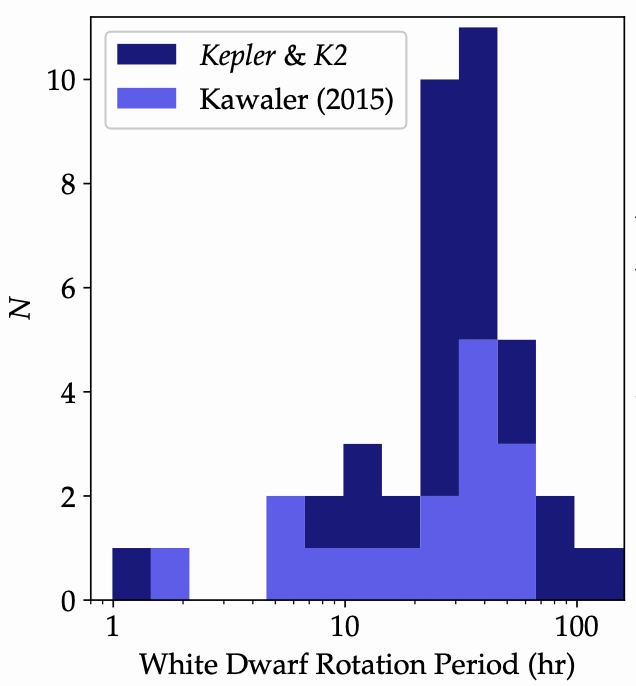}
    \caption{Figure 8 from \citet{2017ApJS..232...23H} showing the distribution in rotation rates measured for 27 pulsating white dwarfs.}
    \label{fig:WD-rot}
\end{figure}

More recent constraints on white dwarf rotation have been made using asteroseismology. For example, \citet{2017ApJS..232...23H} constructed a catalogue of asteroseismic rotation rates for 27 pulsating white dwarfs. These stars fall into a different class of oscillator compared to solar-likes, experiencing non-radial g-mode pulsations. While the class of pulsations is different, much of the same methodology can be used to recover rotation rates from the oscillations. The distribution recovered in \citet{2017ApJS..232...23H} is shown here in figure \ref{fig:WD-rot} -- figure 8 in that work. The peak of this distribution is at $\approx$ 1 day, which is again much slower than would be expected according to traditional models of rotational evolution. This is now seen as additional evidence for the missing angular momentum transport problem, as seen in the red giants and discussed in section \ref{sec:RGB-rot}.

\subsubsection{Magnetism}

\begin{figure}
    \centering
    \includegraphics[width=0.95\linewidth]{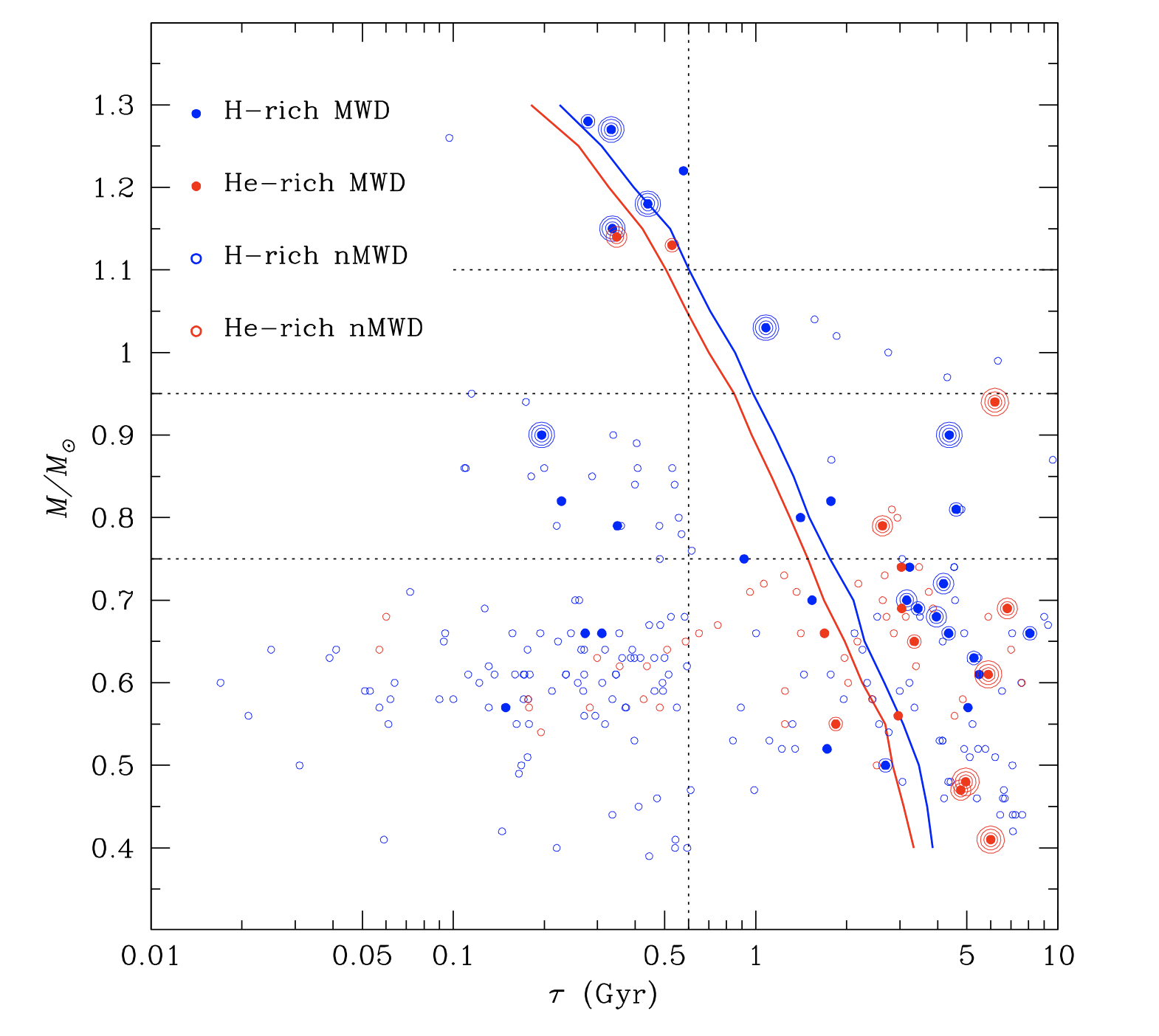}
    \caption{Figure 2 from \citet{2022ApJ...935L..12B} showing the distribution in mass and cooling age of a catalogue of white dwarfs. Those that are measureably magnetic are identified as filled in circles. The strengths of the measured fields are indicated by the concentric circles - more circles means stronger field.}
    \label{fig:WD-mag}
\end{figure}

Magnetic white dwarfs have been observationally identified since the 1970s \citep[see e.g.][]{1970ApJ...161L..77K}. Recent works are now expanding the samples of known magnetic white dwarfs with the aim of understanding the distribution as a collective, and thus perhaps the formation channels that generate this magnetism. An example of this is \citet{2022ApJ...935L..12B}, where-in new spectropolarimetric measurements of magnetic fields were made in 85 white dwarfs, and combined with other measurements from the literature. Figure 2 in that work -- which is included here as figure \ref{fig:WD-mag} -- shows this distribution in mass, cooling age space. In this parameter space, the work finds that the population could be split into two broad categories:
\begin{enumerate}
    \item The most massive white dwarfs, which have strong magnetic field strengths that emerge at the surface almost immediately as the cooling phase commences. These fields appear frequently.
    \item The less massive white dwarfs, where magnetism is much less common. Here fields are weaker, and emerge at the surface slowly and appear to get stronger over time. 
\end{enumerate}
It is noted that this is potentially indicative of two formation channels for magnetism in white dwarfs. For the most massive white dwarfs, it is argued that the observations could be understood if the magnetic fields emerge as a result of a merger event. For the second population, it is argued that this could be explained as the signature of a magnetic field which was generated earlier in the stellar lifecycle, and is slowly relaxing. This field could have been generated in the core convective zone on he main sequence. The observations of core magnetism in red giant stars discussed in section \ref{sec:RGB-mag} could also be consistent with this view. Further works have emerged supporting this multi-formation channel view, including \citet{2024A&A...691L..21C} and \citet{2025ApJ...990...25M}.

Finally, given the connection between observations in red giant cores and what we see in the white dwarf phase, works are beginning to try to piece the two together. A recent example of this is \citet{2026A&A...708L..14E}, which built competing models for magnetism in the stellar interior, and isolated which was simultaneously consistent with observations on the red giant branch and in the lower mass white dwarfs. They find that to do so they have to adopt a magnetic field in the interior that extends beyond the extent of the main sequence core convection zone, which is perhaps evidence that it is not core convection but some other mechanism that produces this field. Their findings can be summarised by figure 2 in that work, which is shown here as figure \ref{fig:WD-mag}.

\begin{figure}
    \centering
    \includegraphics[width=0.95\linewidth]{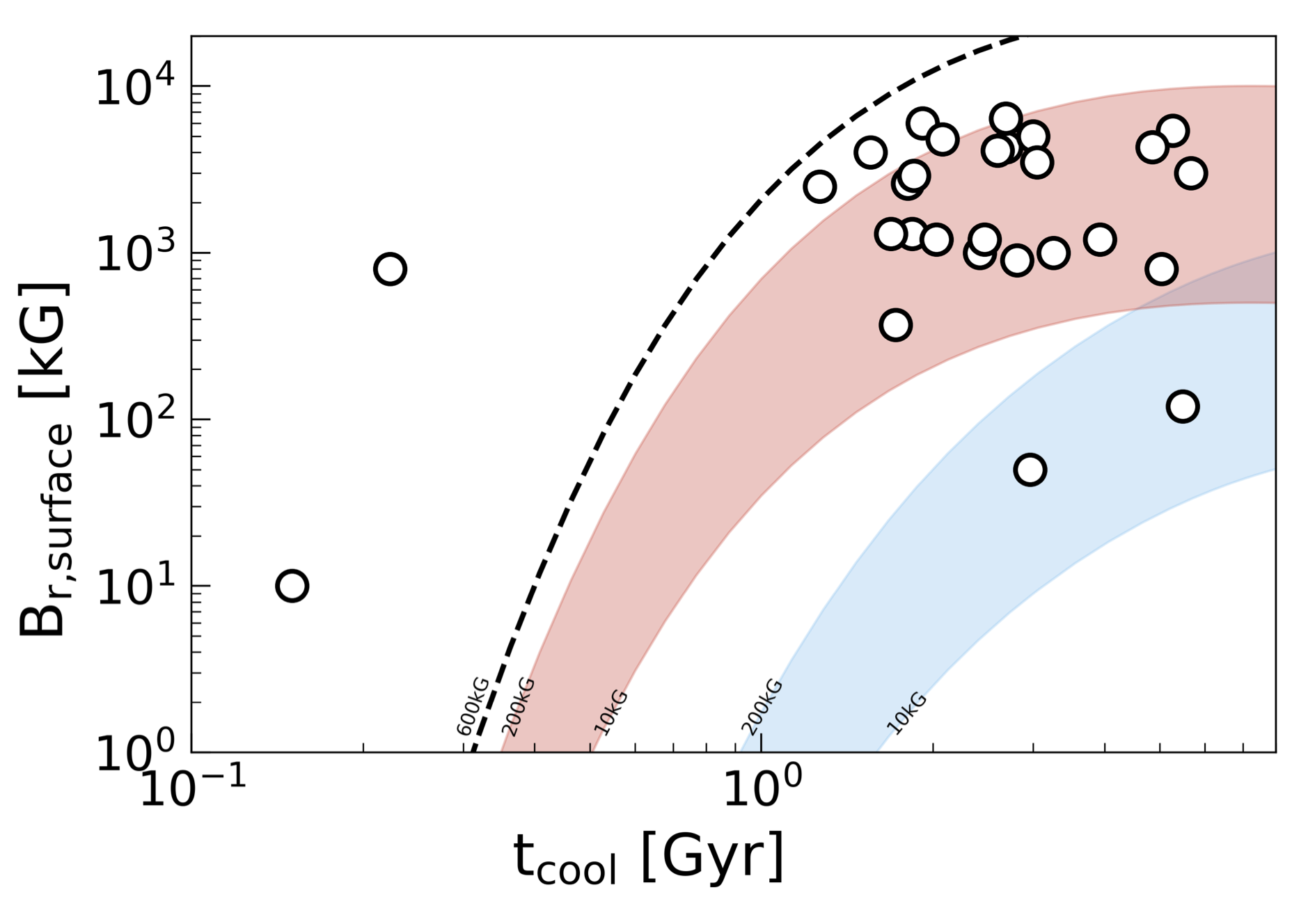}
    \caption{Figure 2 from \citet{2026A&A...708L..14E} showing the average field strength as a function of cooling time for a sample of magnetic white dwarfs. The coloured tracks represent predictions made by models with core field strengths tuned to observations on the red giant branch according to two competing models. In red the magnetic field occupies the entire radiative interior. In blue the field is constrained to the region that experienced core convection on the main sequence.}
    \label{fig:WD-mag}
\end{figure}

\section{Summary}
\begin{figure}
    \centering
    \includegraphics[width=0.9\linewidth]{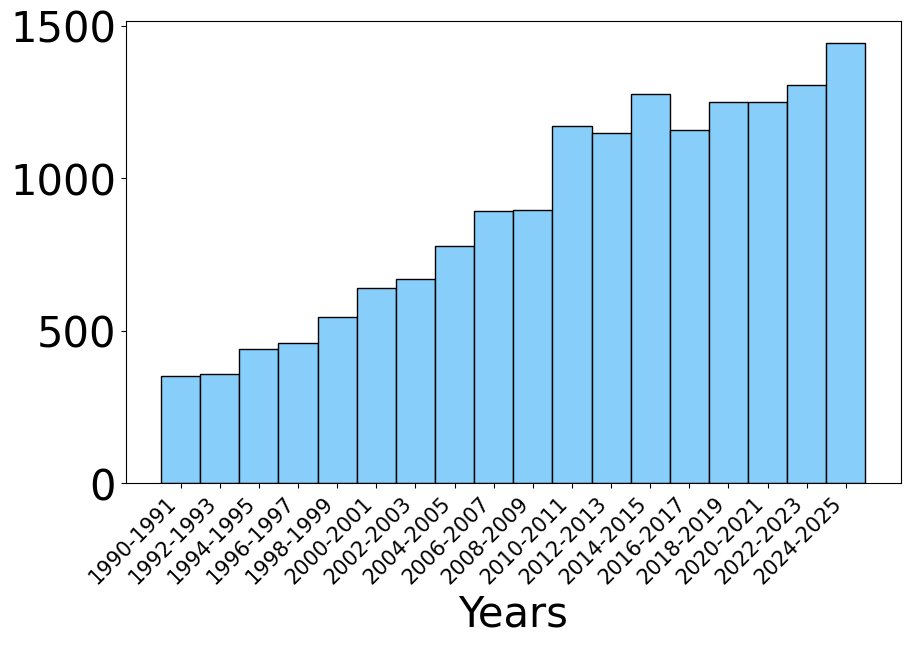}
    \caption{The number of refereed papers published with the keywords rotation, magnetism and stars and listed on NASA's Astrophysics Data System from the years 1990 through to 2025.}
    \label{fig:cite-count}
\end{figure}

While this review has covered a wide range of science in the field of post main-sequence rotation and magnetism, it is far from complete summary of all the work that has happened in the last few decades. Personally, I have noted a significant rise in the number of publications on the topic in just my own field, asteroseismology. Indeed, the question `Have you considered the impact of rotation and magnetism
on your work?' has now transitioned from a point of discussion after talks at conferences, to the focus of entire conference sessions. This is clearly not exclusive to asteroseismology -- For example, figure \ref{fig:cite-count} shows the number of refereed papers with keywords rotation, magnetism and stars published as a function of time between 1990 and 2025. In the last 35 years this number has gone from below 500 to just shy of 1500. 

The works culminating in figure \ref{fig:cite-count} build a picture of post main-sequence rotation and magnetism which is complex, and far from completely understood. There are unanswered questions in every phase, from the subgiants to the white dwarfs. My view is that the pressing missing angular momentum transport problem will be a very active field of research for years to come. One of the bottlenecks in making progress on that front is the seemingly contradictory behaviour in the subgiants versus the red giants. Putting a constraint on where on the subgiant phase/RGB we transition between the two regimes would be critical evidence aiding in understanding the physics at play. To that end, the upcoming PLATO mission will provide asteroseismic data for thousands of subgiants \citep{2024A&A...683A..78G}, allowing us to better sample the transition. 

In many ways, we are just starting to enter the epoch of ensemble inference for post main sequence rotation and magnetism. Asteroseismic missions such as \textit{Kepler} have provided measurements of internal rotation in thousands of stars, and we are approaching a hundred stars with direct measurements of core magnetism. Concurrently, high precision spectropolarimetry and photometry are providing statistical views on the prevalence of surface magnetism, revealing it to be present in low to intermediate mass stars all the way to their final evolutionary stages. Given this key evidence, we are now turning our attention towards interpreting this data as a collective. This will require collaboration across both the theoretical and observational domains and between researchers studying stars across the evolutionary lifecycle.

\section*{Acknowledgments}
{E. Hatt gratefully acknowledges support from the European Research Council (ERC) under the Horizon Europe programme (Calcifer; Starting Grant agreement N◦101165631) and the European Research Council (ERC) under the European Union’s Horizon 2020 research and innovation programme (CartographY; grant agreement ID 804752).}

\bibliographystyle{cs23proc}
\bibliography{example.bib}

\end{document}